\documentclass[fleqn,10pt]{wlscirep}
\usepackage[utf8]{inputenc}
\usepackage[T1]{fontenc}
\title{{\it Ab initio} calculation for electronic structure and optical property of tungsten carbide in a TiCN-based cermet for solar thermal applications}

\author[1]{Shota Hayakawa} 
\author[1,+]{Toshiharu Chono} 
\author[2,+]{Kosuke Watanabe} 
\author[1,2,+]{Shoya Kawano} 
\author[1,2,+,*]{Kazuma Nakamura} 
\author[1,2,3,+,*]{Koji Miyazaki} 
\affil[1]{Graduate School of Engineering, Kyushu Institute of Technology, 1-1 Sensui-cho, Tobata-ku, Kitakyushu, 804-8550, Fukuoka, Japan}
\affil[2]{Integrated Research Center for Energy and Environment Advanced Technology, Kyushu Institute of Technology, 1-1 Sensui-cho, Tobata-ku, Kitakyushu, 804-8550, Fukuoka, Japan}
\affil[3]{Graduate School of Engineering, Kyushu University, 744 Motooka, Nishi-ku, Fukuoka, 819-0395, Fukuoka, Japan}
\affil[*]{kazuma@mns.kyutech.ac.jp; miyazaki.koji055@mail.kyutech.jp} 
\affil[+]{these authors contributed equally to this work}

\begin{abstract}
We present an {\it ab initio} calculation to understand electronic structures and optical properties of a tungsten carbide WC being a major component of a TiCN-based cermet. We found that the WC has a fairly low-energy plasma excitation $\sim$0.6 eV (2 $\mu m$) and therefore can be a good constituent of a solar selective absorber. The evaluated figure of merit for photothermal conversion is prominently high compared to those of the other materials included in the TiCN-based cermet. The imaginary part of the dielectric function is considerably small around the zero point of the real part of the dielectric function, corresponding to the plasma excitation energy. Therefore, a clear plasma edge appeared, ensuring the high performance of the WC as the solar absorber.
\end{abstract}
\begin{document}

\flushbottom
\maketitle
%
%
\thispagestyle{empty}


\section*{Introduction}


The replacement of fossil fuels to renewable energy sources have been intensively investigated in recent years.  Solar energy has been considered as promising alternative to solve global energy issues due to its abundance~\cite{Cao_2014}. Exploring sustainable and eco-friendly technologies has been considered significant to achieve the practical use of harvesting solar energy.  Photovoltaic conversion is most widespread technology to directly generate electric power from solar power.  On the other hand, cermet-based solar absorbers have been also commercialized to obtain thermal energy from the sunlight~\cite{Zhang_1992}. 

\begin{figure}[ht] 
\begin{center} 
\includegraphics[width=0.7\textwidth]{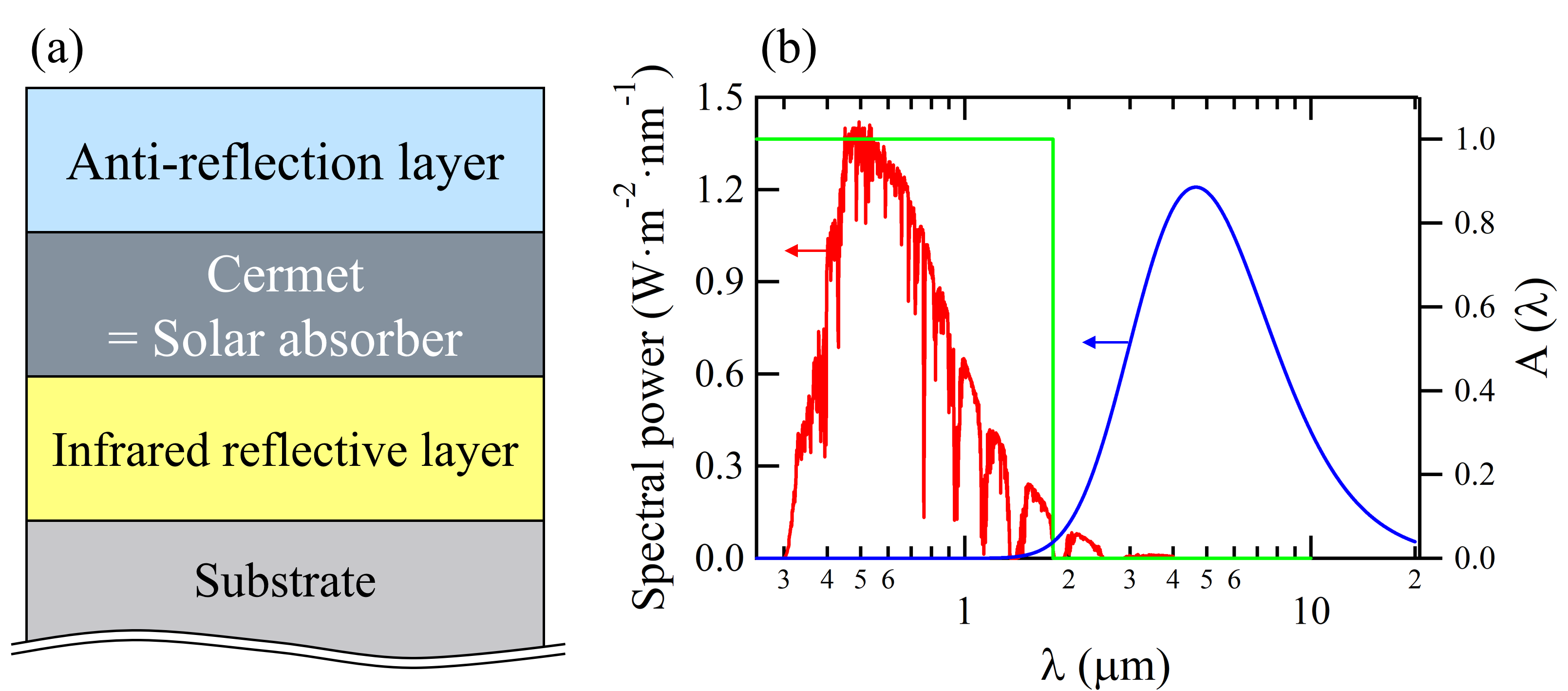}
\end{center} 
\caption{(a) A schematic figure of cermet-based solar absorber consisting of anti-reflection layer, cermet layer, and infrared reflection layer. (b) A schematic figure to see the optical performance required as the solar absorber. A red solid curve describes the spectral power of the sunlight ranging from 0.3 to 2.0 $\mu$m, and a blue solid curve is the spectral power of the thermal irradiation of the black body. Thus, 2 $\mu m$ is a cutoff wavelength ensuring high solar absorptivity and low thermal emissivity. Ideally, a material with an absorptivity spectrum $A(\lambda)$ described with a green solid line is preferable as the solar absorber.}
\label{Solar_absorber}
\end{figure}

Cermet is a composite of metal and ceramic with hardness, thermal stability, and anti-oxidation properties. A typical solar absorber is shown in Fig.~\ref{Solar_absorber}a~\cite{Cao_2014, Zhang_1992}. The cermet-based solar absorber consists of a cermet layer as absorber with an anti-reflection layer on the top and an infrared reflector at the bottom. The solar selective absorber is an important role to achieve the high performance as a solar absorber. The blackbody emissive power significantly increases in high temperature, resulting in large radiative heat loss from the absorber.  An ideal solar selective absorber should have high solar absorptivity and low thermal emissivity as described with a green line in Fig.~\ref{Solar_absorber}b with 2.0 $\mu$m cutoff wavelength~\cite{Sakurai_2014}. The cermet-based absorbers have been well investigated with oxides with metal particles. The dielectric function of the composite is controlled by increasing metal volume fraction to decrease the frequencies of absorption peaks analyzed by Bruggeman approximation~\cite{Zhang_1999}.  

Materials with low-plasma excitation is preferable for solar selective absorber. For further improvement of the performance, the multi-layered solar selective absorbers have been investigated using optical interference. The vacuum deposition process with high production costs is usually needed to precisely control the film thickness~\cite{Zhang_1992, Zhang_1999}. Here, we intend to find materials applicable to a solar selective absorber with low production costs. A TiCN cermet including various metals, carbide, and nitride is generally used for hard machining tools due to their toughness mentioned above~\cite{Arifin_2020}. After the service time by wear of the machining tools, the fine powder of cermet is made as industrial waste disposal. The production costs can be significantly reduced by using wasted materials for solar energy utilization.

In the present study, to discuss the performance of the TiCN-based cermet as a solar selective absorber, we perform {\it ab initio} optical analyses for this cermet, focusing on its major components of W, WC, TiC, and TiN. Systematic investigations of {\it ab initio} optical calculations have been performed for TiN and TiC in a face-centered cubic (fcc) structure~\cite{Kumar_2016,ZHANG_2019} and transition metals including W in a body-centered cubic (bcc) structure~\cite{Romaniello_2006}, but no such calculations have been performed for WC in a hexagonal closed pack (hcp) structure. We will show that the WC has considerably sharp plasma edge around 0.6 eV (2 $\mu m$), which is a highly desirable aspect for the solar absorber. We will discuss the microscopic origin of this low-energy plasma excitation in terms of an {\it ab initio} dielectric analysis. 

The rest of the paper is organized as follows: In "Materials and methods", we specify major components of a TiCN-based cermet to be analyzed. Based on the x-ray diffraction (XRD) pattern analysis, we chose four materials W, WC, TiC, and TiN. We also describe methodological details for {\it ab initio} band-structure and optical-response calculations. "Results and Discussions" describes computational results on the electronic structure, reflectivity spectra, and dielectric functions. We also estimate the figure of merit for photothermal conversion of the four materials. Finally, we describe in "Conclusions" summary of the paper.

\section*{Materials and methods}
\subsection*{TiCN-based cermet}\label{sec:cermet}
\begin{figure}[htpb] 
\begin{center} 
\includegraphics[width=0.5\textwidth]{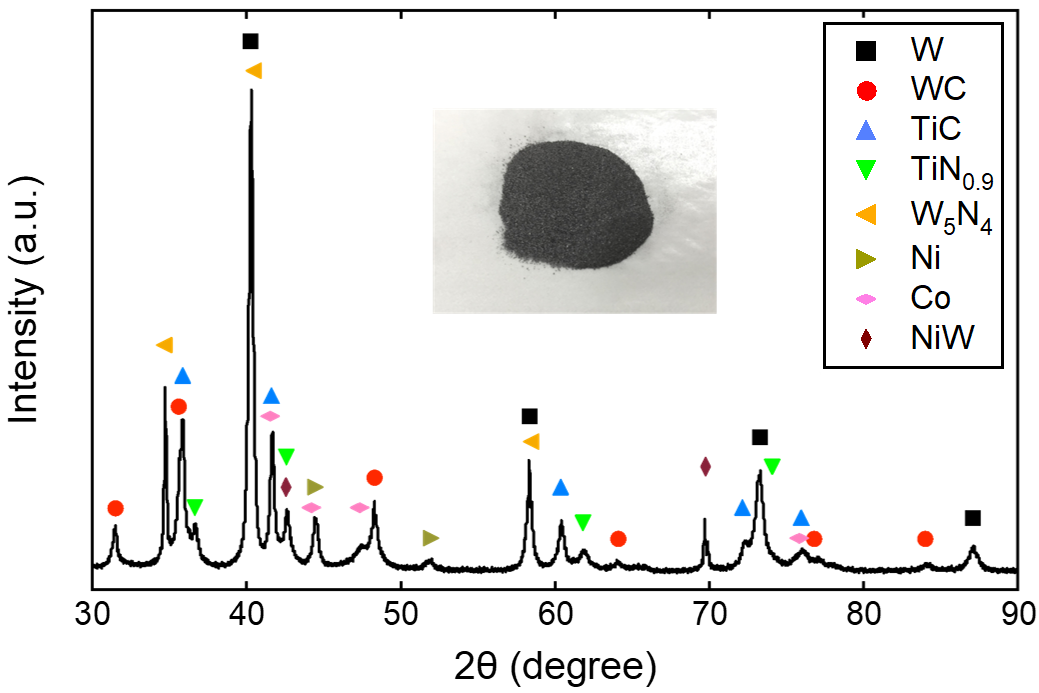}
\end{center} 
\caption{Typical XRD patterns for the TiCN-based cermet as the waste material and the crystal phase identified from the powder-diffraction-file database in the International Centre for Diffraction Data. We see that it contain various components such as transition and rare metals and their carbides and nitrides. In the present study, we focus on major four components W, WC, TiC, and TiN.}
\label{XRD}
\end{figure}

In this section, we describe a component analysis of the scrapped TiCN cermet. It is well known that the cermet consists of various metal and compounds. Our target TiCN-based cermet also contains many components, and each material can individually contribute optical property of the cermet in total. We show in the inset of Fig.~\ref{XRD} the TiCN cermet of the waste material~\cite{PENG201378} to be analyzed. It is black in the visual range, so it can absorb visible light. To specify major components of the TiCN-based cermet,  powder XRD analysis of the cermet powder was carried out by a SmartLab diffractometer (Rigaku Co., Ltd.) with Cu K$\alpha$ radiation ($\lambda$ = 1.5418 \text{\AA}) at the scanning rate of 10.4$^\circ$/min for the 2$\theta$ value of 30 to 90$^\circ$. The result is shown in Fig.~\ref{XRD}. We see that TiC and TiN can clearly be main components, but the present cermet also contain more major ingredients such as tungsten W and tungsten carbide WC. In the present study, therefore, we chose top major four materials WC, W, TiC, and TiN, and calculate their electronic structures and optical properties. We note in passing that all these materials have very high melting temperature near 3000 K (Table~\ref{lattice-constant_bulk}) and are therefore well tolerated to working temperature range in the solar absorber.
\begin{table}[ht] 
\caption{Calculated and experimental lattice parameters of WC, W, TiC, and TiN, where the WC is an hcp structure, and the W is a bcc structure, and the TiC and TiN are an fcc structure. The unit of the lattice parameter is \AA. We also list the melting temperature of each material, which is given in K. }
\begin{center} 
\begin{tabular}{|c@{\ \ \ }|c@{\ \ }c@{\ \ }c@{\ \ } |c@{\ \ }c@{\ \ }c@{\ \ }| c@{\ \ }c@{\ \ }c@{\ \ } |c@{\ \ }c@{\ \ }c|}  \hline 
     & \multicolumn{2}{c|}{WC (hcp)}                            & \multicolumn{2}{c|}{W (bcc)}                             
     & \multicolumn{2}{c|}{TiC (fcc)}                           & \multicolumn{2}{c|}{TiN (fcc)}                           \\ \hline 
     & \multicolumn{1}{c|}{Theory} & \multicolumn{1}{c|}{Expt.} & \multicolumn{1}{c|}{Theory} & \multicolumn{1}{c|}{Expt.} 
     & \multicolumn{1}{c|}{Theory} & \multicolumn{1}{c|}{Expt.} & \multicolumn{1}{c|}{Theory} & \multicolumn{1}{c|}{Expt.} \\ \hline 
$a$  & \multicolumn{1}{c|}{2.922}  & \multicolumn{1}{c|}{2.91 (Ref.~\citen{PhysRevB.38.9483})} 
     & \multicolumn{1}{c|}{3.184} & \multicolumn{1}{c|}{3.165 (Ref.~\citen{lassner1999properties})} 
     & \multicolumn{1}{c|}{4.333} & \multicolumn{1}{c|}{4.328 (Ref.~\citen{karlsson1982optical})}
     & \multicolumn{1}{c|}{4.247} & \multicolumn{1}{c|}{4.250 (Ref.~\citen{karlsson1982optical})} \\ \hline 
$c$  & \multicolumn{1}{c|}{2.847} & \multicolumn{1}{c|}{2.84 (Ref.~\citen{PhysRevB.38.9483})} 
     & \multicolumn{1}{c|}{-    } & \multicolumn{1}{c|}{-                                   } 
     & \multicolumn{1}{c|}{-    } & \multicolumn{1}{c|}{-                                   }
     & \multicolumn{1}{c|}{-    } & \multicolumn{1}{c|}{-                                   } \\ \hline 
$T_{{\rm m}}$ & \multicolumn{2}{c|}{3028 (Ref.~\citen{kurlov2006tungsten})} & \multicolumn{2}{c|}{3149 (Ref.~\citen{calvo2016manufacturing})} & \multicolumn{2}{c|}{2792 (Ref.~\citen{doi:https://doi.org/10.1002/9783527618217.ch16})} & \multicolumn{2}{c|}{2657 (Ref.~\citen{doi:https://doi.org/10.1002/9783527618217.ch16})}  \\   
\hline 
\end{tabular}
\end{center}
\label{lattice-constant_bulk} 
\end{table}

\subsection*{First-principles calculation}\label{sec:method}
To analyze electronic structures and optical properties, we performed {\it ab initio} density functional calculations for the four materials WC, W, TiN, and TiC selected in "TiCN-based cermet" by using Quantum Espresso package~\cite{giannozzi2020quantum}. We used the Perdew-Burke-Ernzerhof type~\cite{perdew1996phys} for the exchange-correlation functional, and the norm-conserving pseudopotentials are generated by the code ONCVPSP (Optimized Norm-Conserving Vanderbilt PSeudopotential)~\cite{hamann2013optimized} and are obtained from the PseudoDojo~\cite{van2018pseudodojo}. We used a 32$\times$32$\times$32 Monkhorst-Pack k-mesh for the Brillouin zone integration. The kinetic energy cutoff is set to be 96 Ry for the wave functions and 384 Ry for the charge density. The Fermi energy was estimated with the Gaussian smearing techniques with the width of 0.001 Ry~\cite{methfessel1989high}. The crystal structures were fully optimized, where the WC is an hcp structure, the W is a bcc structure, and the others TiC and TiN are an fcc structure. The resulting lattice parameters are listed in Table~\ref{lattice-constant_bulk} and are in good agreement with the experimental results~\cite{PhysRevB.38.9483, lassner1999properties, karlsson1982optical}.

{\it Ab initio} calculations for maximally localized Wannier function~\cite{Marzari_1997,Souza_2001} and optical properties were performed with RESPACK~\cite{nakamura2016ab, nakamura2021respack}. For the Wannier function analysis of the WC, TiC, and TiN compounds, we constructed the Wannier orbitals for the $s$ and $p$ orbitals of C and N, and the $d$ orbitals of Ti and W, reproducing the original Kohn-Sham band structures in the energy range from $-$18 eV to 10 eV with reference to the Fermi level. For the bulk W, we constructed the Wannier functions for W-$s$, W-$p$, and W-$d$ orbitals. We performed a decomposition analysis for electronic density of states (DOS) with the resulting Wannier functions. Optical calculations were performed as follows: The energy cutoff for the dielectric function is set to 10 Ry. The total number of bands used in the polarization calculation is 36 for WC, 56 for W, 32 for TiC, and 34 for TiN, which covers unoccupied states up to $\sim$40 eV above the Fermi level. The integral over the Brillouin zone was calculated with the generalized tetrahedron technique~\cite{fujiwara2003generalization} with a smearing of 0.01 eV. For the WC, we performed band and optical calculations considering the spin-orbit coupling, but the obtained results hardly changed (see "Spin-orbit interaction effects in WC")~\cite{PhysRevB.95.195165}. Therefore, in the present discussion, we basically analyze the results based on calculations that do not consider the spin-orbit coupling.

\section*{Results and discussions}\label{sec:res}
\subsection*{Electronic structure}
\begin{figure}[htp] 
\begin{center} 
\includegraphics[width=0.6\textwidth]{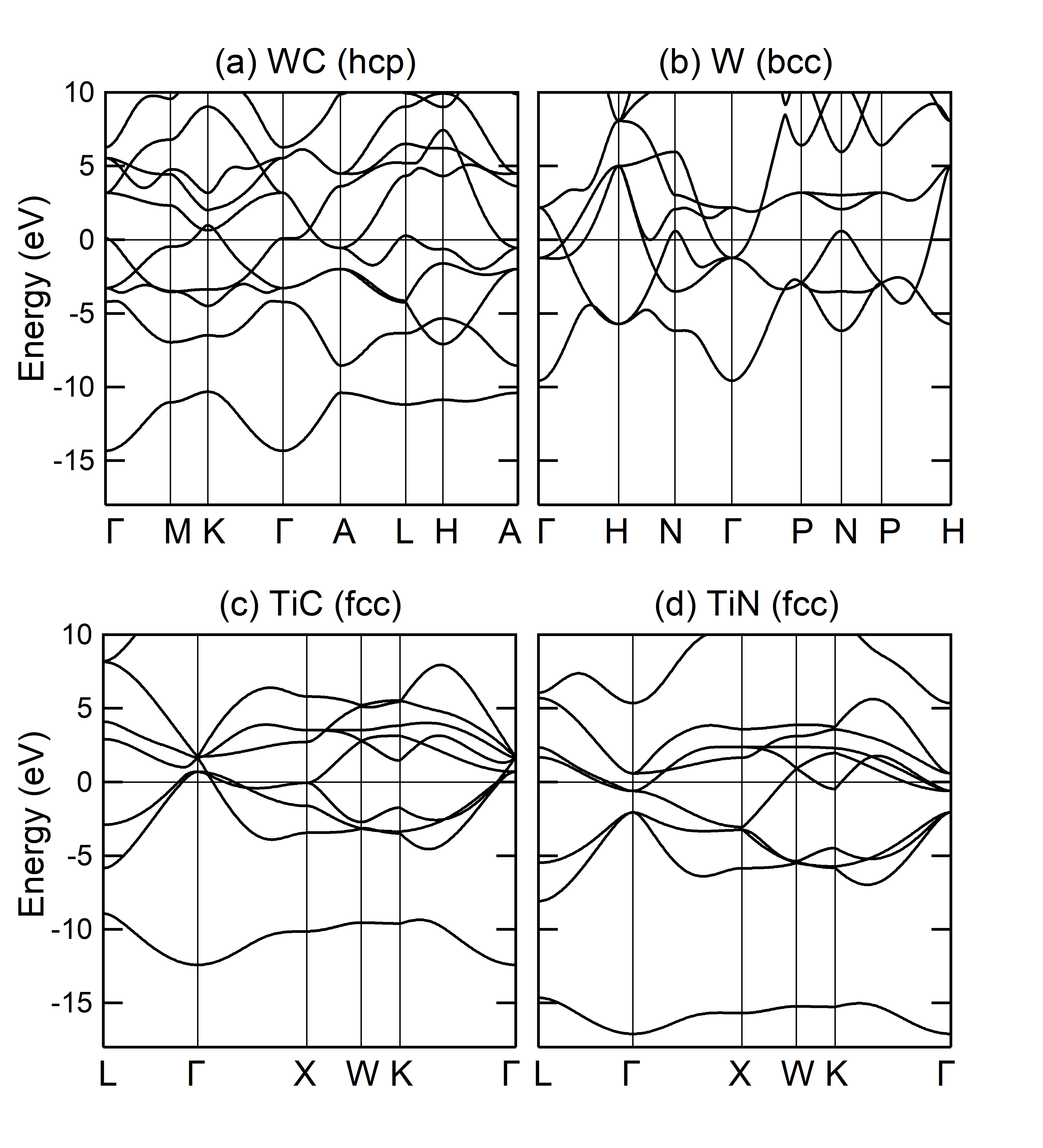} 
\end{center} 
\caption{{\it Ab initio} density functional band structure of (a) WC, (b) W, (c) TiC, and (d) TiN. The energy zero is the Fermi level. Dispersions of the WC are plotted along the high symmetry points in the Brillouin zone, where $\Gamma$= (0, 0, 0), M = (1/2, 0, 0), K = (1/3, 1/3, 0), A = (0, 0, 1/2), L = ($-$1/2, 0, 1/2), and H = (1/3, 1/3, 1/2), where the coordinates are represented in terms of basic vectors of the reciprocal lattice of the hcp lattice. In the band dispersion of the W, H = (1/2, $-$1/2, 1/2), N = (0, 0, 1/2), P = (1/4, 1/4, 1/4), and these coordinates are represented in the basic vectors of the bcc reciprocal lattice. Finally, in the band dispersions of TiC and TiN, L = (0, 1/2, 1/2), X = (1/2, 0, 1/2), W = (1/2, 1/4, 3/4), K = (3/8, 3/8, 3/4), which are based on the basic vectors of the fcc reciprocal lattice.}
\label{band}
\end{figure}

\begin{figure}[htp] 
\begin{center} 
\includegraphics[width=0.78\textwidth]{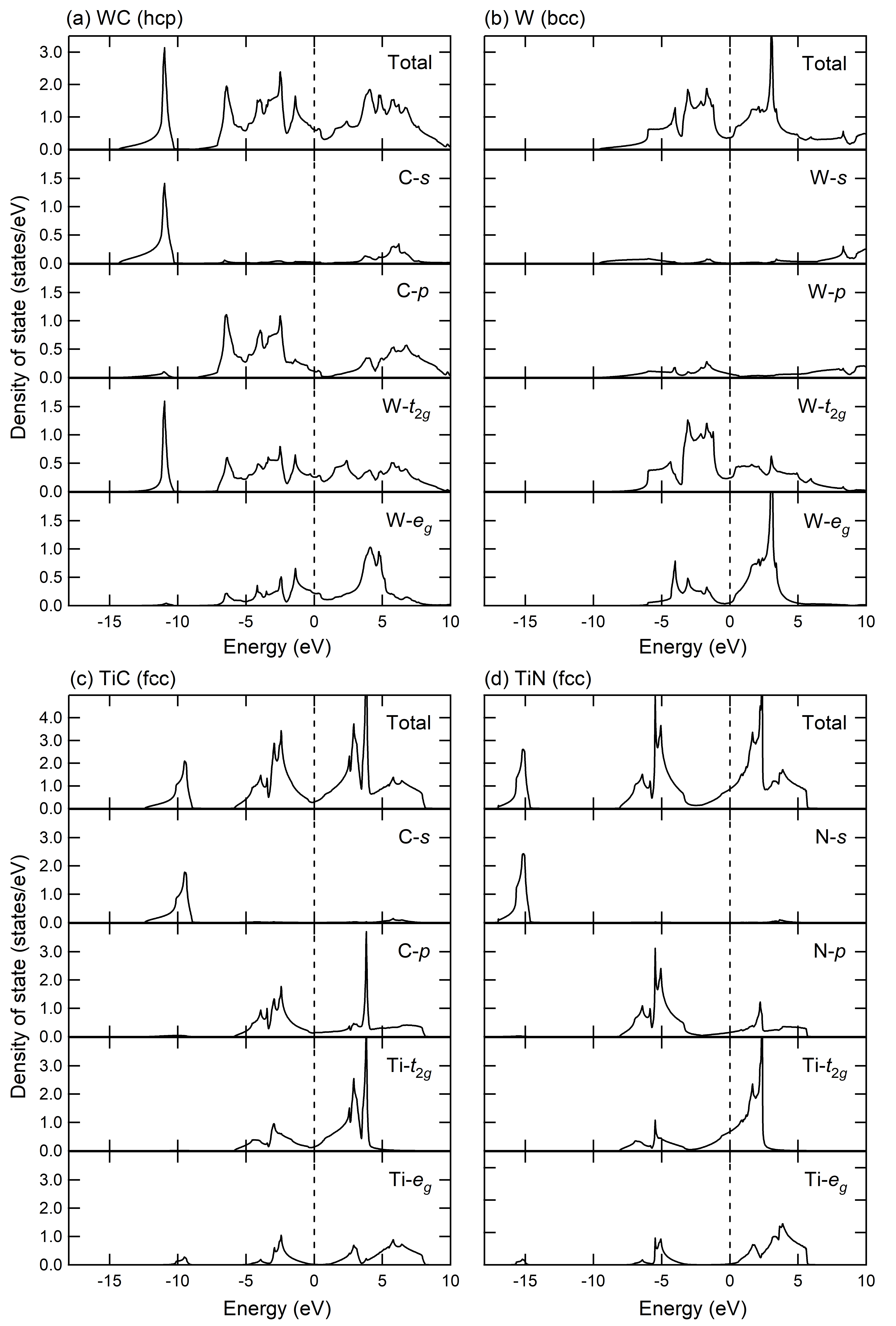}
\end{center} 
\vspace{0.5cm} 
\caption{Calculated DOS of (a) WC, (b) W, (c) TiC, and (d) TiN, and the decomposition into each atomic contribution. We note that these calculations are performed with the maximally localized Wannier functions~\cite{Marzari_1997,Souza_2001}. The energy zero is the Fermi level denoted by a dashed line. For the transition metals W and Ti, we further decompose the contribution into $t_{2g}$ and $e_g$ orbitals.} 
\label{dos-WC}
\end{figure}

Figure~\ref{band} compares our calculated density functional band structures of WC (Fig.~\ref{band}a), W (Fig.~\ref{band}b), TiC (Fig.~\ref{band}c) , and TiN (Fig.~\ref{band}d), where we see that all the materials are metal. Our results are in good agreements with the previous density functional calculations~\cite{suetin2009structural,christensen1974volume,ahuja1996structural}.
We show in Fig.~\ref{dos-WC} our calculated DOS of WC (Fig.~\ref{dos-WC}a), W (Fig.~\ref{dos-WC}b), TiC (Fig.~\ref{dos-WC}c), and TiN (Fig.~\ref{dos-WC}d) and those decomposition into atomic contribution based on the Wannier function, where, for transition metals W and Ti, the local-$d$ DOS is decomposed to the $t_{2g}$ and $e_g$ orbital contributions. For the WC, we see that the W-$d$ and C-$p$ orbitals are well hybridized, while in the TiN, the Ti-$d$ and N-$p$ orbitals are not strongly hybridized, indicating an enhancement of ionic character of the Ti-N bond. Basically, such a bonding tendency can be understand from the view of electronegativity of each atom; W (1.7), Ti (1.54), C (2.55), and N (3.04), taken from Ref.~\citen{Pauling1932THENO}.


\subsection*{Optical property}
\begin{figure}[htp] 
\begin{center} 
\includegraphics[width=0.48\textwidth]{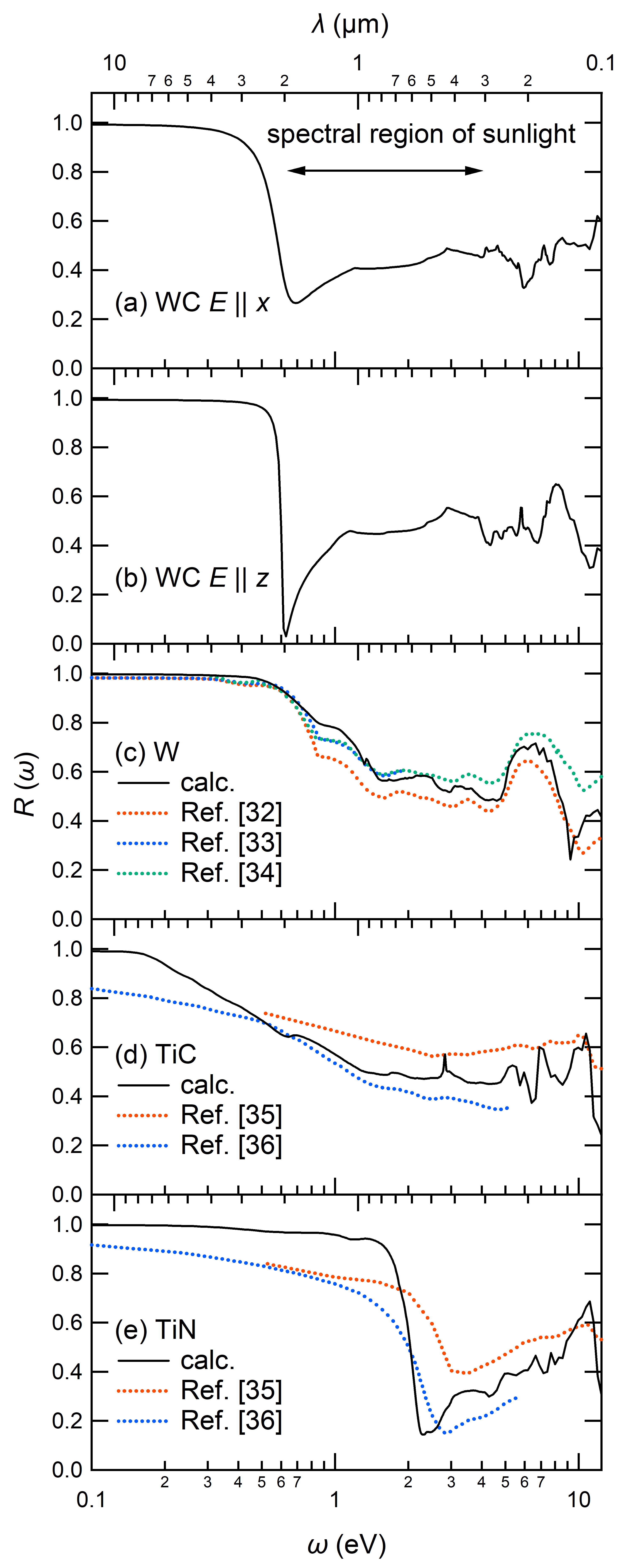}
\end{center} 
\caption{Calculated reflectivity spectra of (a) WC ($E\parallel x$), (b) WC ($E\parallel z$), (c) W, (d) TiC, and (e) TiN as a function of photon energy $\omega$ or photon wavelength $\lambda$ (upper scale). The theoretical curves are given by black solid curves and are compared with the experimental results (colored dotted curves) for W \cite{LYNCH1997275, ordal1988optical, weaver1975optical}, TiC \cite{PFLUGER1997293, roux1982optical}, and TiN \cite{PFLUGER1997293,roux1982optical}. The spectral region of sunlight (0.3$\sim$2.0 $\mu$m or 0.6$\sim$4.1 eV) is indicated with a double arrow in the panel (a).}
\label{reflectance}
\end{figure}

To study optical property of the four materials, we calculated their reflectivity spectra as 
\begin{eqnarray}
 R(\omega) = \Biggl| \frac{ 1-\sqrt{\epsilon(\omega)} }{ 1+\sqrt{\epsilon(\omega)} } \Biggr|^2,
\label{eq:reflectance} 
\end{eqnarray}
where $\epsilon(\omega)$ is a dielectric function in the random phase approximation (RPA) based on the Lindhard formula~\cite{DRAXL_2006}. Figure~\ref{reflectance} compares calculated reflectivity spectra of the WC, W, TiC, and TiN. For the WC, we see a clear plasma edge near 0.6 eV (2 $\mu m$) (Figs.~\ref{reflectance}a, b), and this energy just corresponds to the cutoff wavelength mentioned in Fig.~\ref{Solar_absorber}b. For the TiN (Fig.~\ref{reflectance}e), we also see a sharp plasma edge, but its energy is rather high as 2 eV (0.6 $\mu m$). For other compounds W and TiC, the reflectivity gradually decreases with the frequency $\omega$. We note that the theoretical reflectivity spectra are in a reasonable agreement with the experiment for the W~\cite{LYNCH1997275, ordal1988optical, weaver1975optical}, TiC~\cite{PFLUGER1997293,roux1982optical} and TiN~\cite{PFLUGER1997293,roux1982optical}. To our knowledge, however, there are no experimental data on the reflectivity spectrum of the WC with the hcp structure.

To understand details of the calculated reflectivity spectra, we perform a decomposed analysis for the dielectric function $\epsilon(\omega)$ as  
\begin{eqnarray}
 \epsilon(\omega)=\epsilon_{intra}(\omega)+\epsilon_{inter}(\omega),
 \label{eps_rpa}
\end{eqnarray}
where $\epsilon_{intra}(\omega)$ is the Drude term which is described as  
\begin{eqnarray}
 \epsilon_{intra}(\omega)=1-\frac{\omega_0}{\omega-i\delta}
\label{eps_intra}
\end{eqnarray}
with $\omega_0$ and $\delta$ being the bare plasma frequency and smearing factor, respectively~\cite{nakamura2016ab}. The $\epsilon_{intra}(\omega)$ describes the dielectric response due to the intraband excitation at the Fermi level, and the $\epsilon_{inter}(\omega)$ represents the dielectric response involving the interband excitation. We now write the $\epsilon(\omega)$ with introducing switching parameters $s$ and $t$ as 
\begin{eqnarray}
 \epsilon(\omega)= 1-s\frac{\omega_0}{\omega-i\delta}+t\epsilon_{inter}(\omega), 
 \label{eps_s_t}
\end{eqnarray}
where the $s=t=1$ case describes the full RPA dielectric function in Eq.~(\ref{eps_rpa}). On the other hand, with the setting of ($s=1$, $t=0$), the dielectric function considers only the Drude term [Eq.~(\ref{eps_intra})], while the setting of ($s=0$, $t=1$) describes the dielectric function with only the interband excitation. By comparing the dielectric functions under the various parameter setting, we discuss details of the dielectric functions. In the practical calculations, by taking the inverse of the dielectric matrix in the plane-wave basis, we calculate the macroscopic dielectric functions with ${\bf G}={\bf G}^{\prime}={\bf 0}$ and the ${\bf q}\to{\bf 0}$ limit~\cite{nakamura2021respack}, where ${\bf G}$ and ${\bf G}^{\prime}$ are reciprocal lattice vectors, and ${\bf q}$ is the wave vector in the Brillouin zone. 

In the discussion of the dielectric function, we focus on the two quantity; one is the plasma frequency $\omega_p$ characterized as the zero point of the dielectric function, and the other is the plasmon-scattering strength $W_p$ due to the interband excitations, estimated from the imaginary part of the dielectric function at $\omega=\omega_p$. For better solar absorber, it is desirable that the $\omega_p$ is near the cutoff energy 0.6 eV (or the cutoff wavelength 2 $\mu m)$, and around there, the $W_p$ should be small.

\begin{figure}[htp] 
\begin{center} 
\includegraphics[width=0.45\textwidth]{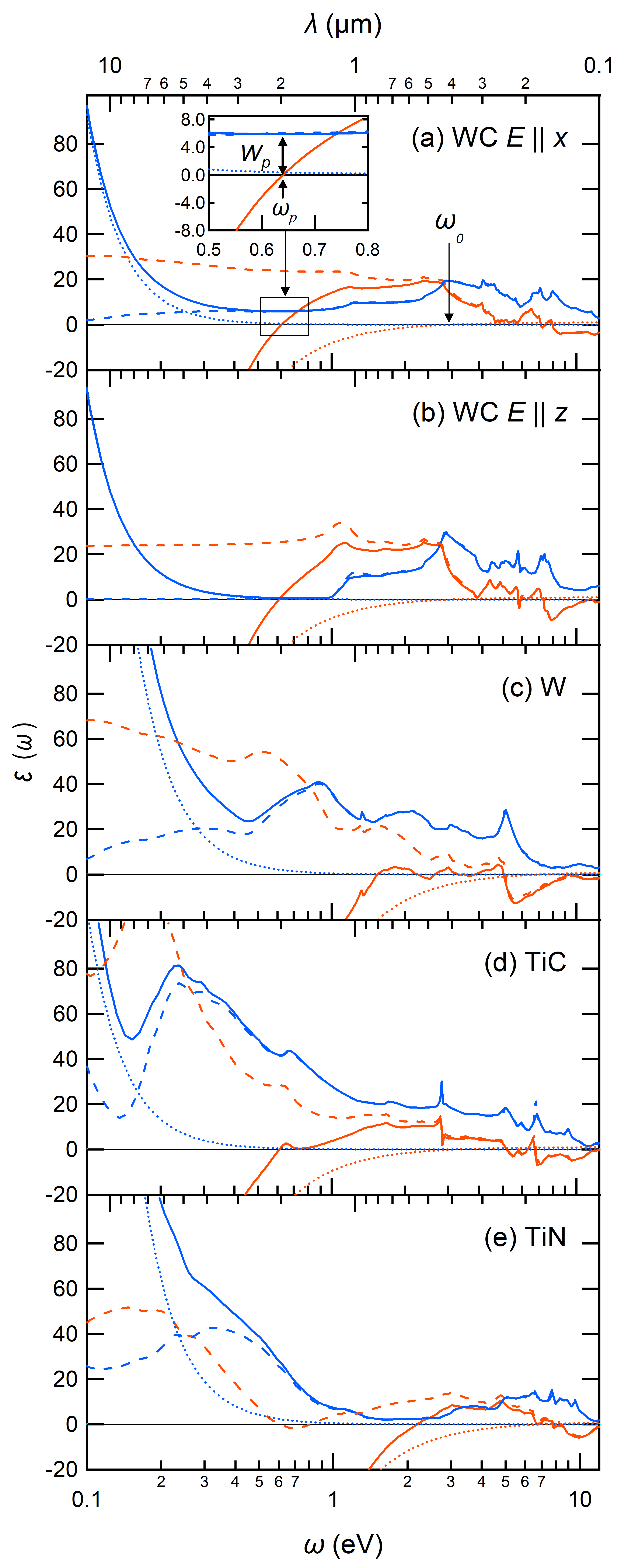}
\end{center} 
\caption{Calculated dielectric function of (a) WC ($E\parallel x$), (b) WC ($E\parallel z$), (c) W, (d) TiC, and (e) TiN. Red and blue solid curves describe the real and imaginary parts of the dielectric function in Eq.~(\ref{eps_rpa}), respectively. Red and blue dotted curves represent the real and imaginary parts of the Drude dielectric function in Eq.~(\ref{eps_intra}), respectively. The dashed red and blue curves are the real and imaginary parts of the dielectric function without the Drude term [i.e., $s=0$ and $t=1$ in Eq.~(\ref{eps_s_t})], including only the contribution from the interband transitions. An inset in the panel (a) shows an enlarged view in the energy range [0.5 eV: 0.8 eV], in which the plasma frequency $\omega_p$ and plasmon scattering $W_p$ are indicated. The bare plasma frequency $\omega_0$ indicated by an arrow is specified from the Drude dielectric function.}
\label{dielecric_function}
\end{figure}

Figure~\ref{dielecric_function} compares {\it ab initio} dielectric functions of the four materials. The solid red and blue curves are the real and imaginary parts of the full macroscopic dielectric function [Eq.~(\ref{eps_rpa}) or $s=t=1$ in Eq.~(\ref{eps_s_t})], respectively. The dotted red and blue curves respectively describe the real and imaginary parts of the macroscopic dielectric function only considering the Drude term [Eq.~(\ref{eps_intra}) or $s=1$ and $t=0$ in Eq.~(\ref{eps_s_t})]. The dashed red and blue curves respectively represent the real and imaginary parts of the macroscopic dielectric function with only considering the interband transitions [$s=0$ and $t=1$ in Eq.~(\ref{eps_s_t})]. Through the comparison, we find the several aspects, and let us consider the case of the WC as an example (Fig.~\ref{dielecric_function}a):
\begin{enumerate}
\item With neglecting the Drude contribution (dashed curves), the resulting dielectric function gives the insulating behavior; the real part of the dielectric function (the dashed red curve) gives the finite value at the limit $\omega\to 0$, and the imaginary part (the dashed blue curve)  goes to zero of this limit. In the case of the WC, the real part is flat around the low-energy region. 
\item By considering the Drude term (solid curves), the dielectric function (the solid red curve) rapidly drops toward minus infinite and therefore the zero point is formed in the low-energy region.
\item Thus, the bare plasma excitation $\omega_0$ ($\sim$ 3 eV) due to the $\epsilon_{intra}(\omega)$, denoted by the arrow in Fig.~\ref{dielecric_function}a, is largely reduced to $\omega_p \sim$ 0.6 eV (the arrow in the inset) by considering the interband transition.
\item This trend is basically common among all the materials. 
\item An interesting point is that, in the case of the WC (Figs.~\ref{dielecric_function}a, b), $W_p$ is appreciably small around the plasma excitation $\omega_p$. Thus, the sharp plasma edge appears in the reflectivity spectra of the WC in Figs.~\ref{reflectance}a, b. 
\end{enumerate}

\subsection*{Spin-orbit interaction effects in WC} \label{SOI} 
\begin{figure}[!tbp] 
\begin{center} 
\includegraphics[width=0.475\textwidth]{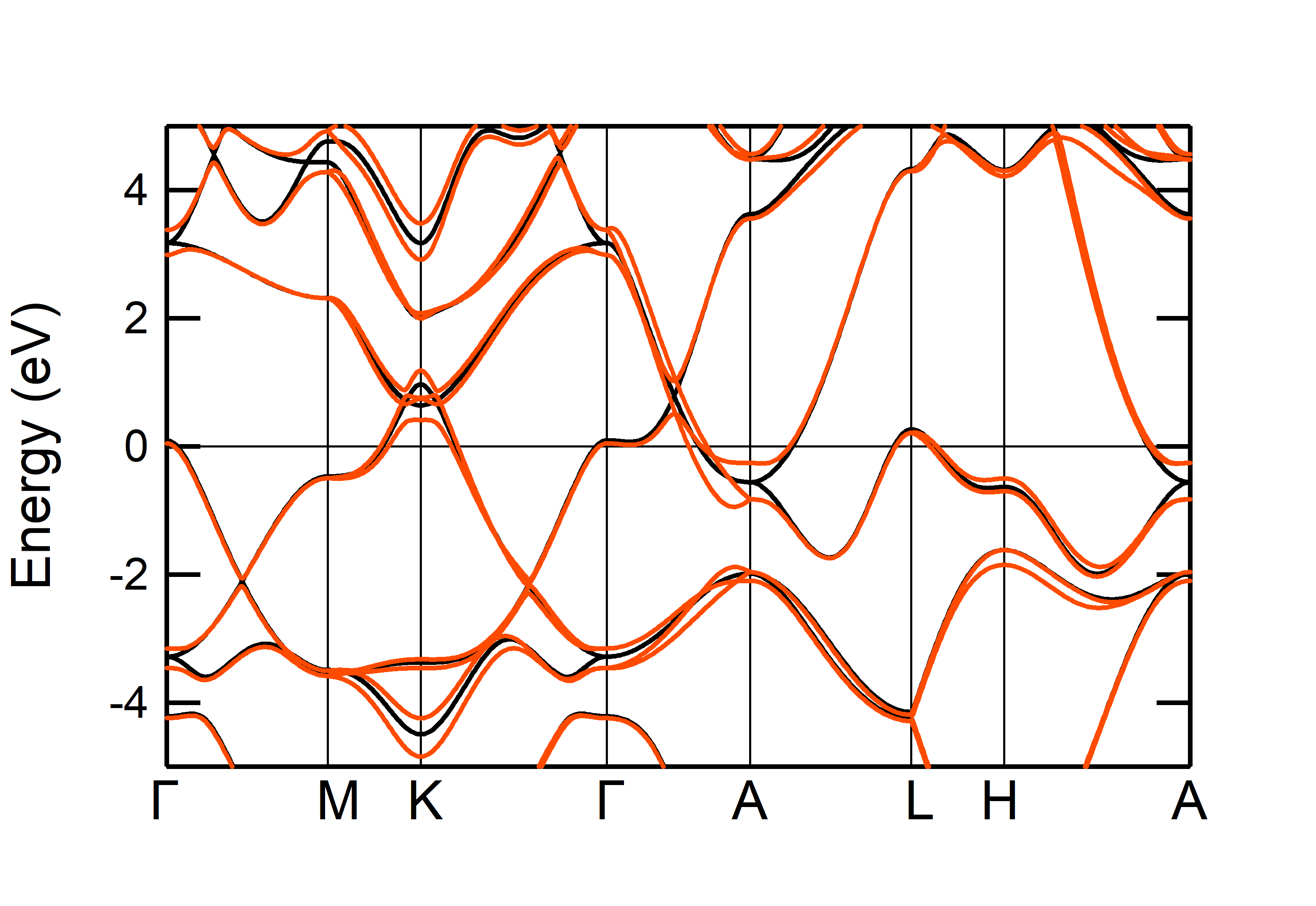}
\end{center} 
\caption{Comparison between {\it ab initio} density functional band structure of WC with (red solid curves) and without (black solid curves) the SOI. Path of the band dispersion is the same as Fig.~\ref{band}a.}
\label{WC-spin-band}
\end{figure}

\begin{figure}[!tbp] 
\begin{center} 
\includegraphics[width=0.475\textwidth]{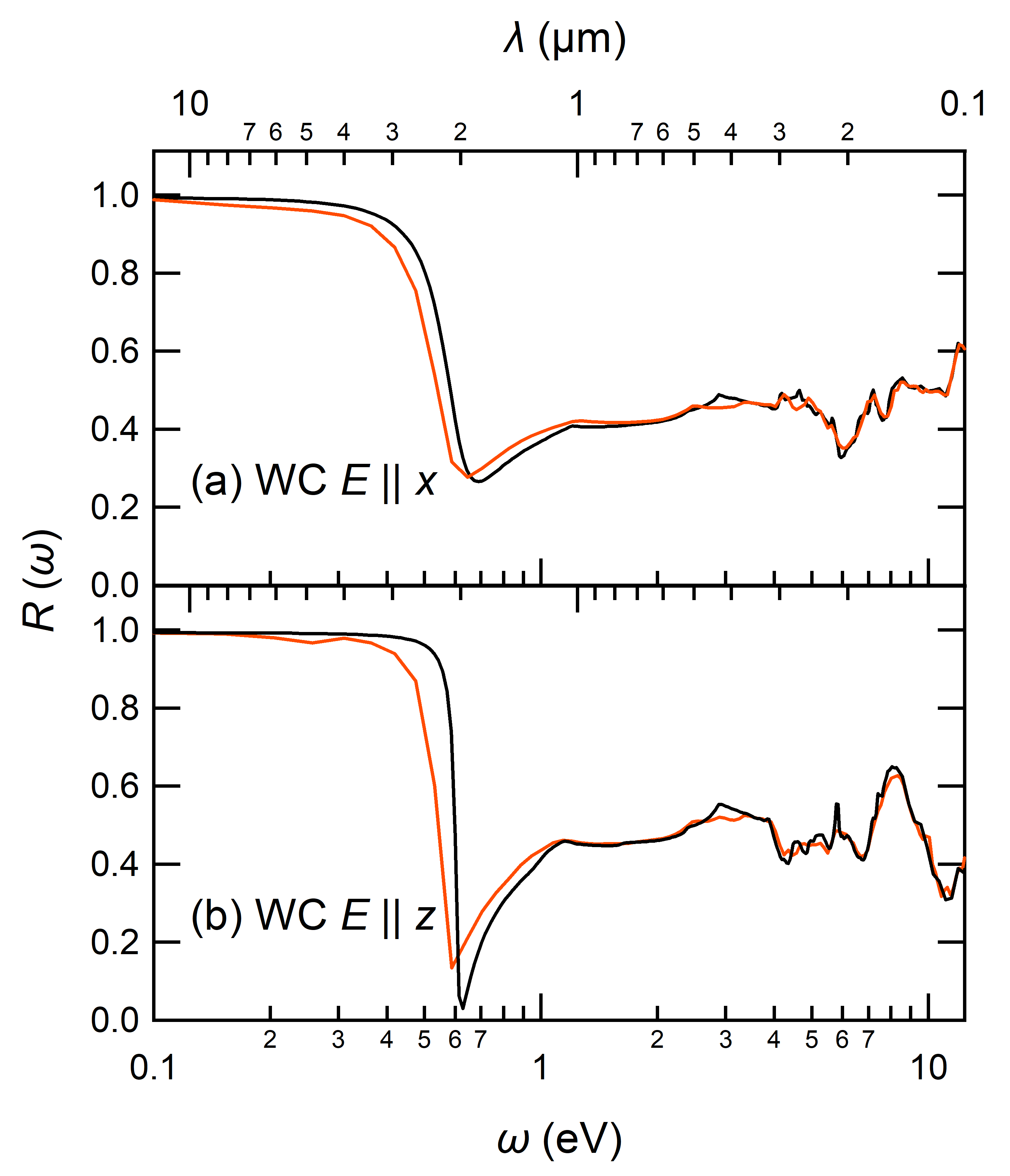}
\end{center} 
\caption{Comparison between {\it ab initio} reflectivity spectra of WC with (red solid curves) and without (black solid curves) the SOI. The panels (a) and (b) show the results of $E\parallel x$ and $E\parallel z$, respectively. The view of the figure is the same as Fig.~\ref{reflectance}.}
\label{WC-spin-reflectivity}
\end{figure}

Here, we discuss a spin-orbit interaction (SOI) effect in WC. The SOI of tungsten is known to be nearly 0.4 eV~\cite{Mattheiss_1964}. Figure~\ref{WC-spin-band} compares our calculated band structures with (red solid curves) and without (black solid curves) the SOI. The SOI can bring about a splitting of the low-energy bands, but the effect is basically small overall. We also compare in Fig.~\ref{WC-spin-reflectivity} our calculated reflectivity spectra, where the red and black solid curves are the results with and without the SOI, respectively. The spectra with the SOI are calculated with the spinor version of RESPACK~\cite{Charlebois_2021}. We again see a small difference between the two results, so we think that the SOI effect can be ignored within the purpose of evaluating the reflectivity or absorptivity performance of the WC. 

\subsection*{Figure of merit for photothermal conversion}
Solar absorptivity and thermal emissivity are widely used to evaluate the performance of solar selective absorbers. The solar absorptivity $\alpha_s$ is defined via the wavelength integral as~\cite{Cao_2014,bermel2012selective,Zhang_2022}
\begin{equation}
\alpha_s = \frac{\int_{\lambda_l}^{\lambda_h} \bigl(1-R(\lambda)\bigr) I_{sol}(\lambda) d\lambda}{ \int_{\lambda_l}^{\lambda_h} I_{sol}(\lambda) d\lambda}, 
\label{equation_alp}
\end{equation}
where $R(\lambda)$ is the reflectivity spectra as a function the wavelength $\lambda$, which is taken from the present {\it ab initio} calculations. $I_{sol}(\lambda)$ is the spectral solar radiance (air mass of 1.5) taken from Ref.~\citen{GUEYMARD2001325}. The $\lambda_l$ and $\lambda_h$ are the lower and higher cutoff wavelengths, respectively, and were set to 0.28 $\mu$m and 4 $\mu$m in the present study. Similarly, the thermal emissivity at a temperature $T$ is defined as follows~\cite{Cao_2014,bermel2012selective,Zhang_2022}: 
\begin{equation}
\epsilon_t(T) = \frac{\int_{\lambda_L}^{\lambda_H} \bigl( 1-R(\lambda) \bigr) I_{b}(T, \lambda) d\lambda}{\int_{\lambda_L}^{\lambda_H} I_{b}(T, \lambda) d\lambda}. 
\label{equation_eps}
\end{equation}
Here $I_b (T,\lambda)$ is the spectral blackbody radiative intensity, which is taken from Ref.~\citen{https://doi.org/10.1002/andp.19013090310}. The $\lambda_L$ and $\lambda_H$ are the lower and higher cutoff wavelengths for the emittance evaluation, respectively, and were set to 0.1 $\mu$m and 124 $\mu$m in the present study. The photothermal conversion efficiency is measured via the following figure of merit (FOM)~\cite{bermel2012selective}:
\begin{equation}
\eta_{{\rm FOM}} = B \alpha_s - \frac {\epsilon_t(T) \sigma T^4 } {c I_0}
\label{equation_eta}
\end{equation}
where $\sigma$, $c$, and $I_0$ are the Stefan–Boltzmann constant, the solar concentration, and the solar flux intensity, respectively. $B$ is related to the transmittance of glass envelope, and is typically chosen to be 0.91 (Ref.~\citen{zhang1999high}). In the present calculation, we set $T$ to 673 K, $c$ to 80 suns, and $I_0$ to 1 kW/m$^2$. These are a standard condition~\cite{doi:10.1021/acs.chemrev.5b00397}.

We summarize in Table~\ref{tb:arrst} our calculated parameters of the present materials, characterizing the performance of the solar absorber. About the plasma frequency $\omega_p$, the WC and TiC are clearly small as 0.6 eV, but the WC plasmon scattering $W_p$ at $\omega=\omega_p$ is appreciably small as 0.7-5.9 compared to the TiC (41.8). Thus, the solar absorptivity $\alpha_s$ of the WC becomes appreciably high compared to the other materials, and yields a better performance of the figure of merit $\eta_{{\rm FOM}}$. On the other hand, we comment that the $\alpha_s$ of WC is still not so high as 0.53-0.57, which can be improved with better choices of anti-reflection and/or infrared reflective layers sandwiching the cermet layer (Fig.~\ref{Solar_absorber}), which is left as an important issue for the future study.
\begin{table}[hbtp]
 \caption{Summary of the bare plasma frequency $\omega_0$, plasma frequency $\omega_p$, the imaginary part of the dielectric function $W_p$ at $\omega=\omega_p$, solar absorptivity $\alpha_s$ in Eq.~(\ref{equation_alp}), thermal emissivity $\epsilon_t$ at $T=673$ K in Eq.~(\ref{equation_eps}), and figure of merit $\eta_{{\rm FOM}}$ for photothermal conversion in Eq.~(\ref{equation_eta}) at $T=673$ K. The $\omega_0$ and $\omega_p$ are given in the unit of eV.} 
  \vspace{.25cm} 
  \centering
  \begin{tabular}{|c@{\ \ \ \ }|c@{\ \ \ \ }|c@{\ \ \ \ }|c@{\ \ \ \ }|c@{\ \ \ \ }|c@{\ \ \ \ }|c|} \hline  
    Material                 & $\omega_0$ & $\omega_p$ & $W_p$ & $\alpha_s$ & $\epsilon_t$ & $\eta_{{\rm FOM}}$ \\ \hline 
    WC ($E\parallel x$) & 3.03 & 0.63 & 5.90 & 0.57 & 0.04 & 0.51  \\ \hline 
    WC ($E\parallel z$) & 3.04 & 0.62 & 0.67 & 0.53 & 0.02 & 0.48  \\ \hline 
    $\rm{W}$                 & 7.15 & 1.53 & 23.9 & 0.33 & 0.01 & 0.30  \\ \hline 
    $\rm{TiC}$               & 3.25 & 0.61 & 41.8 & 0.49 & 0.09 & 0.43  \\ \hline 
    $\rm{TiN}$               & 7.15 & 2.18 & 2.34 & 0.37 & 0.01 & 0.33  \\ \hline
  \end{tabular}
  \label{tb:arrst}
\end{table}

\section*{Conclusion}\label{sec:conclusion}
In the present paper, we have studied electronic structures and optical properties of WC, W, TiC, and TiN, identified as major components in the TiCN-based cermet. We have found that the WC exhibits a sharp plasma edge due to the low-energy plasma excitation $\sim$0.6 eV (2 $\mu m$), which just corresponds to a cutoff wavelength suitable for the solar selective absorber. We have checked that this result is unchanged with taking into account the SOI of W. We have analyzed the origin of the low-energy plasma edge and found that, in the WC, the plasmon scattering due to the interband transitions is strongly suppressed around the plasma excitation. This aspect directly reflects to the solar absorptivity, bringing about the better performance of the figure of merit for photothermal conversion. The solar absorptivity of the WC would further be improved to suppress the reflection due to the interband excitation with the fine-structure processing and/or introduction of reflection layers.

\subsection*{Data Availability} The data that support the findings of this study are available from the corresponding author upon reasonable request. 


\providecommand{\noopsort}[1]{}\providecommand{\singleletter}[1]{#1}%

\subsection*{Acknowledgments} The authors would like to acknowledge MARUWAGIKEN Co., Ltd. for providing a cermet powder preparation. This research was supported by JSPS KAKENHI Grant Number JP19K03673, JP20K05100, JP22H01183, and Fukuoka Research Commercialization Center for Recyclling Systems. 

\subsection*{Author Contributions}
The role in the present paper is as follows: Conceptualization, K.N. and K.M.; software, K.N.; formal analysis, S.H., T.C., K.W., S.K., and K.N.; investigation, S.H., T.C., K.W., S.K., K.N., and K.M.; writing–original draft preparation, S.H., T.C., K. W., K.N., and K.M.; writing–review and editing, K.N. and K.M.; supervision, K.N. and K.M.; project administration, K.N. and K.M.; All authors reviewed the manuscript.

\subsection*{Competing interests} The authors declare no competing interests.

\subsection*{Additional information}
\textbf{Correspondence} and requests for materials should be addressed to K.N. or K.M. 



\end{document}